RESEARCH ARTICLE  OPEN ACCESS

# Andriod Based Punjabi TTS System

Hardeep [1], Parminder Singh [2]
M.Tech. Scholar [1], Associate Professor [2]
Department of Computer Science and Engineering
Guru Nanak Dev Engineering College
Ludhiana, India

**ABSTRACT**
The usage of mobile phones is nearly 3.5 times more than that of personal computers. Android has the largest share among its counter parts like IOS, Windows and Symbian Android applications have a very few restrictions on them. TTS systems on Android are available for many languages but a very few systems of this type are available for Punjabi language. Our research work had the aim to develop an application that will be able to produce synthetic Punjabi speech. The paper examines the methodology used to develop speech synthesis TTS system for the Punjabi content, which is written in Gurmukhi script. For the development of this system, we use concatenative speech synthesis method with phonemes as the basic units of concatenation. Some challenges like application size, processing time, must be considered, while porting this TTS system to resource-limited devices like mobile phones.

## I. INTRODUCTION

In the past years, mobile phones from being just a medium of communication have now become a gadget with features of many other devices clubbed into a single device. They have become such an essential part of our life and a lot of our routine work depends on them.

Android OS has become very popular lately. The Android mobile devices also support the text-to-speech synthesis. Thus, speech synthesis has become an important modality on mobile phones as various aspects such as driving, jogging, screen size, etc. restrict the use of visual modality. This TTS application also helps the users to read written text while jogging, driving etc. Android TTS application is helpful for users with visual disabilities and illiterate masses [1]. Speech synthesis systems for mobile phones are difficult to implement as they have limited storage capacity and computing performance.

## II. SPEECH SYNTHESIS TECHNIQUE

Speech synthesis is the automatic generation of speech waveforms that convert the input text data to speech waveforms. Concatenating the pre-recorded speech that is stored in database produces synthesized speech. There are different techniques to speech synthesis, articulatory modeling, rule-based and concatenative techniques. Nowadays, speech research area uses the Concatenative speech synthesizers. Thus, we have developed our Punjabi synthesizer using the above-mentioned technique. There are numerous barriers in text pre-processing, like, abbreviations, numerals and acronyms.

Concatenative synthesis depends upon stringing together of segments of recorded speech; we used phonemes as the basic unit for reducing both number and size of the database. Furthermore, Concatenative technique does the suitable matching of various phonemes in an acceptable processing time. A TTS system based on three modules [2]:

- Transcription: This module transcribes the orthographic input text into a sequence of phoneme, which specifies the sound to be produced.
- Prosody: This module computes phonemes durations, determine word and sentence-level stress and assign a fundamental frequency contour to the utterance.
- Synthesis: This module synthesizer the desired utterance from the specification provided by the transcription and the prosody module also converts the written text into the speech synthesis. Speech synthesis is the artificial production of the human speech which allows the users to create their own synthetic voice.

The TTS synthesizer is composed of two phases as mentioned in the fig. 1 front end and back end. The two are analysis which converts the input text to a phoneme in the front end. Thereafter, back end converts the phoneme to waveforms that can output as sound [3].

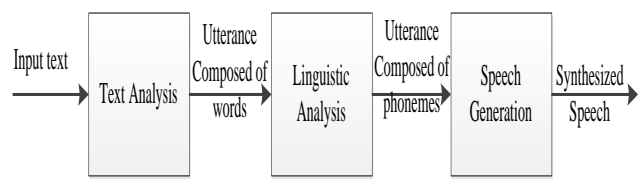

Fig. 1: Text-To-Speech Engine

## III. RELATED WORK

Until now, TTS systems have developed for many platforms, like PC's and mobile devices. However, most applications are for English language. The problems faced in porting a TTS system to mobile devices are limited storage and processing power. Gopi et al. [4] developed Malayalam language TTS synthesizer for Android platform. Authors used an ESNOLA technique for speech generation based on concatenation technique using diaphones like segments as a basic unit for concatenation. Ahlawat and Dahiya developed English and Hindi TTS systems for the Android environment [5]. For English TTS, they used phonemes as the smallest units for concatenation. For Hindi TTS, authors used two layers: first getting the input text in Hindi and, then mapped it with the Entire Hindi data into English language. Kaur and Singh [6] developed a Punjabi speech synthesizer for android mobile devices using a concatenative method, but sound quality needs to be improved using DSP techniques. Singh and Lehal developed a computer based Punjabi TTS system [7] and they used syllables as the basic unit of concatenation, which produces the high quality synthesised sound. Mhamunkar et al. developed a speech recognition system that is an on-line Speech to Text engine using hidden Markov Models (HMM- Hidden Markov Model). The application takes input in the form of voice, then searches the word in mobile dictionary and gives the output as the synthetic speech [8].

Mukherjee and Mandal developed a Bengali TTS synthesis system for the Android devices [9] using ESNOLA (Epoch Synchronous Non Overlap Add) technique based on concatenative speech synthesis technique using Partnemes as the basic units for concatenation.

## IV. ANDROID ARCHITECTURE

Google on 5th November 2007 launched the Android Mobile platform for mobile devices like such as smart phones, net books and PDA [2]. Android is a Linux Kernel based open source OS, which goes about as a reflection layer between the software stack and hardware. All applications written in Java Programming language and Eclipse is the IDE for developing Android apps. Google created Dalvik as the virtual machine environment [10] for mobile devices for compiling the projects, Runs on VM, not on the Java VM. Since embedded systems have the constraints of application size and processing time, so Dalvik Virtual Machine (DVM) is optimized [2] for low memory requirement this is its main feature. The architecture of Android operating system is divided into four layers: the first layer is Application layer, the second layer is Application framework, and the third layer is divided in two sub layers: libraries and Android Runtime, and the last layer is Linux Kernel. Therefore, in total there [10] are five layers. Android OS having the following key features:

- It has free use and manufactures can easily adapt the operating System of mobile devices.
- It is the most efficient way of using memory with DVM.
- The visual and audio content has Very good quality. User can easily develop many application using various tools and valuable database of software libraries

## V. METHODOLOGY

The methodology followed in developing this system is as follows:

(1) The Concatenative synthesis technique is used for development of application to get the naturalness quality of the synthetic speech that takes real recorded speech: marking the second's position and concatenating these sound positions back together during synthesis to produce the required output speech.
(2) The user browses the Punjabi text file from the storage media from either phone memory or SD card. Then, we divide the file into words. This step analyzes the position of consonant and vowel from the word. Thereafter, segment the words into phonemes as a consonant and vowel and their combination.
(3) The framework taking into account the Punjabi speech prepared database that consists of starting seconds and end seconds positions of phoneme. The sound of the phonemes has been marked very precisely from the recorded wave. The database preparation with the selection of all the consonants and vowels and their combination, recording of these sentences performed by professional speaker and finally marking start and end position of phoneme sound in these recorded wave sound files. For the storage of, Punjabi phonemes and their corresponding seconds start and end positions we used SQlite DBMS.
(4) Now the phonemes are searched in the database, retrieve the start, and end second's positions, which are then searched in the application resource for its corresponding sound to be played. At last, these phonemes are concatenated to get the sound desired to the input text.

## VI. DEVELOPMENT OF PUNJABI TTS SYSTEM

The aim of our research is to develop a TTS system that as an output produces the speech of the input text with very less response time on android-based mobile phones. Thus, we used the concatenative speech synthesizer technique to get the two qualities of the desired output: Naturalness and intelligibility. We used Punjabi phonemes as the basic unit. For the development of Punjabi speech database, the Punjabi language containing the valid phonemes with the combination of vowels (V) and consonant-vowel (CV) [11] and their corresponding sound positions are stored in the database file. The second's starting and ending position for V and CV are marked from the recorded wave file. The input text of any length is segmented into Punjabi phonemes, thereafter phonemes are searched in the prepared database and corresponding second's position is retrieved. Thereafter, these second's position search in sound wave file is stored in the assets folder.

### A. Database Preparation

These are the few steps, which are followed for the generation of Punjabi speech database.

1) *Punjabi phoneme selection:* The development for the Punjabi TTS, phoneme was selected as the smallest unit of

concatenation, which was made up of consonants and vowel and their combination. In Punjabi language, there are two types of phonemes V and CV that produces the 380 phonemes with nasalized vowels and 380 phonemes with non-nasalized vowels [12], resulting total 722 valid Punjabi phonemes, where vowel and consonant shown in TABLE I.

The major reason for selection of phonemes as basic speech unit is that, phoneme is smaller than word and syllables. So, phoneme give lass number of speech sound as compare to words and syllables, thus requiring less storage space.

2) *Sentence for Recording:* The total sets of phonemes of Punjabi language are carefully analyzed to get the minimum required phonemes for the system. For labelling the phonemes sounds, we selected the words having all the consonants and vowel and their combination for making the recording.

3) *Word Recording:* The professional female speaker of Punjabi recorded the selected words. The quality of speech depends upon quality of recorded sound and also the quality of extracted speech units from this recorded sound. Recording was done in the studio with the following characteristic: Sample rate: 44,100Hz, Bit Depth: 16 bit, Channels: Mono

4) *Labelling Phonemes Sounds:* The next step is to label the phoneme sound in the recorded wave sound file. This is an exceptionally lengthy process, and should be done carefully. Since the synthesized speech totally depends upon how accurately the phonemes boundaries have been marked. For the process Sonic Foundry Sound Forge 10.0 has been used. After carefully analyzing and listening to the word sounds the phonemes sounds have been marked by noting down the starting and ending second's position of each phoneme.

5) *Punjabi Speech Database:* The database is an important part of a Punjabi TTS based on the concatenation technique. For the development of the Punjabi TTS system, database has been developed. The designed database for the TTS system includes three fields: Phonemes, Starting Second's position and end second's position.

B. *Words Separation into Phonemes*

Phoneme is the basic unit of concatenation, so we initially segment the Punjabi text into words and then the words into phonemes, which are already stored in database. To epitomize, the word "ਹਰਦੀਪ" will be segmented into five phonemes: ਹ (ਹ ਅ) + ਰ (ਰ ਅ) + ਦੀ (ਦ ਈ) + ਪ (ਪ ਅ).

TABLE II
VALID AND INVALID PHONEMES

| Phoneme Type | Number of Phonemes | Invalid Phonemes | Valid Phonemes | Example |
|---|---|---|---|---|
| V (Non-Nasalized) | 10 | 0 | 10 | ਜਾ |
| V (Nasalized) | 10 | 0 | 10 | ਜਾਂ |
| CV (Non-Nasalized) | 380 | 7 | 373 | ਦੀ [ ਦ(C) + ਈ(V)] |
| CV (Nasalized) | 380 | 51 | 329 | ਮਾਂ [ ਮ(C) + ਆਂ(V)] |
| Total | 760 | 58 | 722 | |

C. *Finding Phonemes and concatenation*

In this step, we search the corresponding phonemes in database and retrieve the sound positions. Thereafter, we search the desired sound position in the application resource and store it in the memory. For the particular phoneme, if the search is successful then the corresponding sound position will be retrieved from database, else it will be skipped as invalid phoneme if no such kind of entry exists. At last, we concatenate these phoneme sounds to get the sound according to the input text.

D. *Porting TTS to Android Platform*

The final step of our TTS application is to port to the Android platform from Computer system. 'Gurmukhi' font has been stored in the application to run the browsed Punjabi text. By using Gurmukhi font, for every Punjabi written text from the browsed file, its related Punjabi letter is displayed in the text field. The text then converted into the Unicode value and these Unicode values are the input for the TTS engine.

## VII. PROPOSED SYSTEM ARCHITECTURE

The minimum requirement specification to implementing the TTS on Android OS are version 2.2 and 512MB RAM. It can run easily due to the limited processing power, constrained battery life and storage capacity. The size of this synthesizer is 23MB; after the sound once get played then it will close the database connection and release the memory. The application has the functionality that the user generates Punjabi speech sound by browsing the Punjabi text file from the storage area after clicking on the 'CLICK HERE TO SELECT FILE'. Then, the application displays the written text, when the user clicks on 'start Reading' it generates the speech file according to the written text and plays the audio file. The application can also read inbox messages for the user. For this, the user will click on 'CLICK HERE TO SELECT SMS', and it will display all the inbox messages and the user can then choose the Punjabi message. Then, after clicking on 'start reading', it will produce the sound. Fig 2 illustrates the Interface of our TTS system and Fig 3 illustrates the Speech Synthesis system.

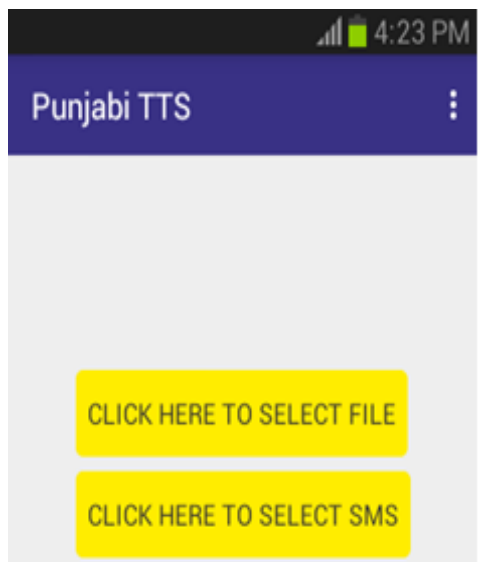

Fig. 1  Interface of TTS application

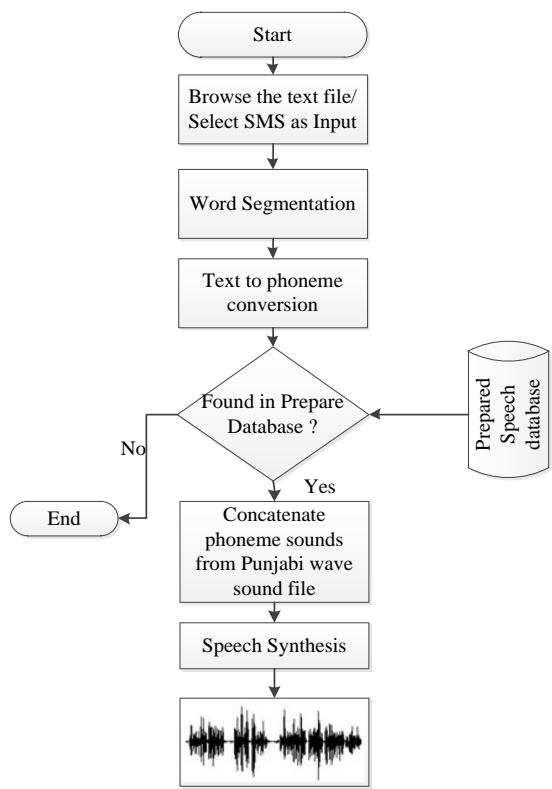

Fig. 2  Flow chart of Punjabi TTS

## VIII.  CONCLUSIONS

In this paper, a Punjabi speech synthesizer has been developed for Android mobile devices. The aim of this application was to develop a TTS system that as an output produces the speech of the input text with very less response time on android based mobile phones. The idea behind the development of this system is the concatenation of phonemes, using phonemes as a basic speech unit.  The size of this Punjabi TTS application is 23 MB. The application shows desirable results for the segmentation of words into phonemes. This application segments the text of any length into phonemes. The proposed system was developed keeping in mind the needs of visually impaired people and for the people who cannot read and write as it will read out the text for the user.